\documentclass[11pt]{elsart}

\usepackage{graphicx}
\usepackage{epsf}
\usepackage{wrapfig}
\usepackage{here}
\usepackage{amssymb}

\begin{document}

\begin{frontmatter}

\title{Monte Carlo studies on the sensitivity of the HEGRA imaging
atmospheric \v{C}erenkov telescope system in observations of extended
$\gamma$-ray sources}
%[Monte Carlo studies on the sensitivity of the HEGRA IACT system]

\author[1]{A. Konopelko},
\author[1,2]{F. Lucarelli},
\author[1]{H. Lampeitl} and
\author[1]{W. Hofmann}

\bigskip
\address[1]
{Max-Planck-Institut f\"ur Kernphysik, D-69029
Heidelberg, Germany}
\address[2]
{Facultad de Ciencias Fisicas, Universidad Complutense,
E-28040 Madrid, Spain}

\begin{abstract}
In this paper, we present the results of Monte Carlo simulations of
atmospheric showers induced by diffuse $\gamma$-rays as detected by
the high-energy gamma-ray astronomy (HEGRA) system of five imaging
atmospheric \v{C}erenkov telescopes (IACTs). We have investigated the
sensitivity of observations on extended $\gamma$-ray emission over
the entire field of view of the instrument. We discuss a technique to
search for extended $\gamma$-ray sources within the field of view of
the instrument. We give estimates for HEGRA sensitivity of
observations on extended TeV $\gamma$-ray sources.

\vspace{3 mm}



\end{abstract}
\end{frontmatter}

\section{Introduction}
During the last decade, very-high-energy (VHE) $\gamma$-ray astronomy
($E_\gamma > 100$~GeV), which utilizes ground-based imaging
atmospheric \v{C}erenkov telescopes (IACTs), has made a substantial
contribution to the $\gamma$-ray astrophysics of a number of
extra-galactic and galactic objects (for a review, see
\cite{rene_ong,weeks_catanese}). One of the reasons has been the
tremendous progress in the observational technique. One can point out
two major trends in this direction. The first is the use of imaging
cameras with very fine pixellation (a pixel size of about
0.1$^\circ$), equipped with fast electronics and an intelligent
trigger, for a single stand-alone telescope, accomplished by Whipple
\cite{whipple1}, CAT \cite{CAT1} and CANGAROO \cite{cangaroo1}.
Secondly, there has been the development of the stereoscopic
observational technique with a number of IACTs with imaging cameras
of relatively coarse pixellation (a pixel size of 0.25$^\circ$),
primarily by the high-energy gamma-ray astronomy (HEGRA)
collaboration \cite{HEGRA1}. Both trends have finally converged in
three future experiments---H.E.S.S. \cite{hess1}, CANGAROO III
\cite{cang4} and VERITAS \cite{veritas1}, which are the systems of a
10~m class of telescope. Another major project, called MAGIC
\cite{magic1}, is a single telescope with a very large reflector of
17~m. Aspects of stereoscopic observations with three such telescopes
have been discussed in \cite{duo}.

For $\gamma$-ray point sources, the sensitivity of the imaging
atmospheric \v{C}erenkov technique substantially relies on the
angular resolution of the instrument. Note that the methods of
cosmic-ray rejection based on the analysis of image shape are still
not effective enough to reduce entirely the background of hadronic
air showers. Thus, in addition, a good angular resolution
significantly reduces the background contamination induced by the
isotropic cosmic rays.

The observations of $\gamma$-ray showers with a single telescope do
not allow a complete geometrical reconstruction of the shower axis in
space, because only one projection of a shower is available per
event. A full reconstruction becomes possible in observations with
two or more telescopes offering simultaneously a number of views of
an individual shower. However, using advanced methods based on the
strong correlation between the shape of the \v{C}erenkov light images
and their angular distances to the source position, one can achieve
for a single IACT an angular resolution for an individual
$\gamma$-ray shower of about $0.12^\circ$
\cite{ulrich,lebohec1998,lessard2001} using a fine pixellation
camera.

A substantial improvement of the angular resolution has been achieved
by the HEGRA collaboration using a stereoscopic system of five IACTs
with rather coarse camera pixellation. The stereoscopic observations
with such a system allow us to reach, with good quality data, an
angular resolution for an individual $\gamma$-ray as good as
0.06$^\circ$ \cite{hofmann1}. The rich HEGRA data sample of
$\gamma$-rays from the Crab~Nebula \cite{hegra_crab}, Mkn~501
\cite{hegra_mkn501} and Mkn~421 \cite{hegra_mkn421} have allowed us
to prove in great detail this angular resolution, which is in good
agreement with the predictions based on the Monte Carlo simulations
(for details, see \cite{HEGRA1}). Such an angular resolution has
allowed us to perform a systematic search for point-like $\gamma$-ray
sources at the flux level of $10^{-11}\, \rm erg\, cm^{-2} s^{-1}$.
Such high sensitivity was confirmed by the detection of a very faint
$\gamma$-ray source, the supernova remnant (SNR) Cas~A, which
steadily emits $\gamma$-rays at the flux level of about $5.9\times
10^{-13}\, \rm ph \, cm^{-2} \, s^{-1}$ above 1~TeV
\cite{hegra_casa}.

The HEGRA system of five IACTs has proved a very effective tool to
search for TeV $\gamma$-ray emission and to study the energy spectra
of point-like sources, which are well established in observations at
other wavelengths. However, the potential $\gamma$-ray source may
appear anywhere within the entire field of view (FoV) of the
instrument, or it may have a rather large angular size as compared to
the angular resolution of the instrument, for a certain number of
important tasks, such as: (i) to search for $\gamma$-ray emitters
with poorly known position (such as the EGRET unidentified sources);
ii) to perform sky surveys; (iii) to study extended $\gamma$-ray
emission from supernova remnants; (iv) to investigate diffuse
emission from the Galactic plane; and finally, (v) to detect the
primordial $\gamma$-ray bursts. In the following, we investigate the
sensitivity in detecting $\gamma$-ray emission with the HEGRA system
of IACTs in observations of this type. For this purpose, we have
performed Monte Carlo simulations of diffuse $\gamma$-rays as well as
isotropic cosmic rays (see sections~2 and 3). The simulations are
compared with real data taken with the HEGRA system of IACTs
(section~4). An important issue is the efficiency of applying the
orientational and shape cuts in order to distinguish the
$\gamma$-rays from the background isotropic cosmic rays. Here we
study how this efficiency depends on the angular distance of a
$\gamma$-ray source to the centre of the FoV (section~5). In
section~6, we discuss the analysis techniques to search for
point-like and extended $\gamma$-ray sources with the HEGRA system of
IACTs, and finally in section~7 we give sensitivity estimates for
shell-type supernova remnants.

\section{Simulations}
The simulations have been carried out using a two-step procedure,
which has been described in detail in \cite{HEGRA1}. In the first
step, the showers of primary $\gamma$-rays and protons were generated
in the atmosphere using the ALTAI Monte Carlo code \cite{altai}. The
showers were randomized over an impact distance with respect to the
centre of the telescope array up to a maximum of 250~m. The major
parameters of the samples of the simulated showers are summarized in
table~1. The simulations have been carried out for different zenith
angles, covering the range of angles usually used in observations of
extended sources with the HEGRA system of IACTs. In the second step,
the response of the telescope camera was simulated for all previously
generated showers. This procedure accounts for all efficiencies of
the \v{C}erenkov light detectors (for details, see \cite{marc}). For
the protons as well as for the $\gamma$-rays, an additional
randomization of the images within a solid angle around the telescope
axis with a half opening angle of $4.0^\circ$ was introduced in order
to simulate the isotropic flux of the proton showers and the diffuse
$\gamma$-rays.

The simulations were carried out for the final HEGRA set-up of five
IACTs\footnote{As the commissioning of the telescope system extended
over several years, early data sets use three or four telescopes.}.
Each of the telescopes consists of an 8.5~m$^2$ reflector focusing
the \v{C}erenkov light on to a camera of 271 photo-multipliers
arranged in a hexagonal matrix covering a FoV with an effective
radius of 2.15$^\circ$. The telescope camera triggers when the signal
in two next-neighbours of the 271 photo-multiplier tubes exceeds a
threshold of 8~mV ($\simeq 6$~photoelectrons), and the system readout
starts when at least two telescopes are triggered by \v{C}erenkov
light from an atmospheric shower. The simulation procedure briefly
described above was previously used for the production of the
simulated data for the HEGRA system of IACTs \cite{HEGRA1}. Different
parameters of the simulated cosmic-ray-induced as well as
$\gamma$-ray-induced atmospheric showers have been compared to the
data (see \cite{HEGRA1,marc1,hofmann2,hofmann3}).

% Table 1 -------------------------------------------------------

\begin{table}[t]
\caption{Summary of the sample of Monte Carlo simulated showers.}
\vspace{3 mm}
\begin{center}
\begin{tabular}{llll}\hline
Primary  & Zenith angle ($^\circ$) & No of events & Energy (TeV) \\ \hline
$\gamma$ & 0          & 33239      & 0.1 - 30 \\
         & 20         & 106810     & 0.4 - 30 \\
         & 30         & 96729      & 0.4 - 30 \\
         & 45         & 74641      & 0.4 - 30 \\ \hline
p        & 0          & 21119      & 0.2 - 50 \\
         & 20         & 30153      & 0.2 - 50 \\
         & 30         & 41717      & 0.3 - 100 \\
         & 45         & 29028      & 0.3 - 100 \\ \hline
\end{tabular}
\end{center}
\end{table}
% ---------------------------------------------------------------

\section{Analysis}
The HEGRA system of IACTs is the first stereoscopic system to perform
routine observations. Its basic idea is to observe atmospheric
showers from a number of different viewing angles. The stereoscopic
analysis of the system data is based on the geometrical
reconstruction of the shower development in space, and consequently
allows us to measure the arrival direction of each individual shower
(for details, see \cite{hofmann1}). By superposition of the several
images in one common focal plane, one can determine the intersection
point of the major axes of the ellipsoid-like images. This
intersection point gives the direction of the shower. When the
telescope system points directly towards the object, the
reconstructed source position is in the centre of the FoV. However,
in most observations of point-like $\gamma$-ray sources with the
HEGRA system of IACTs, so-called `wobble mode' observations have been
made. In such a case, the position of the source is offset by
$0.5^\circ$ from the camera centre, allowing us to perform continuous
ON-source observations with the corresponding OFF region taken at
1$^\circ$ offset from the source position across the camera centre.

In observations of extended $\gamma$-ray sources or diffuse
$\gamma$-ray emission, one can divide the FoV into square bins of a
size approximately equal to the angular resolution, accumulate
ON-source and OFF-source events in these bins according to the
reconstructed shower directions, and calculate the significance of a
possible excess for each bin. Events are pre-selected according to
their image shapes, to reduce the cosmic-ray background. Such an
analysis is discussed below in detail. An alternative analysis
technique is to use a grid which is much finer than the resolution,
and to accumulate for each grid point the ON-source and OFF-source
events with the directions consistent with a source localization at
that point \cite{gal_scan}. This technique is suitable to search for
isolated point sources of an unknown location. But it has the
disadvantage that the significances determined for adjacent grid
points are highly correlated, which makes it difficult to judge the
overall significance for an extended source.

In the case when an appropriate model of diffuse $\gamma$-ray
emission can be easily constructed, one can use the likelihood ratio
method to test the hypotheses (see section~7). If the expected
emission region is relatively small, i.e.~less than 0.5$^\circ$ in
radius, in observations of such sources one can use a wobble mode
approach by setting a corresponding offset of the target from the
centre of the FoV, and by using the angular area symmetric across the
centre of the FoV for the background estimate.

\subsection{Detection rate}
The counting rate of isotropically distributed cosmic rays and
diffuse $\gamma$-rays after the application of shape cuts, $\tilde
R_{(\gamma,cr)} \rm \left[Hz\,\,str^{-1}\right]$, is defined as
\begin{equation}
\tilde R_{(\gamma,cr)}(\Theta) = R_{(\gamma,cr)}(\Theta)\eta_{(\gamma,cr)}(\Theta),
\end{equation}
where $R_{(\gamma,cr)}$ is the counting rate of cosmic rays or
$\gamma$-rays before the cuts, and $\eta_{(\gamma,cr)}$ is the
acceptance after applying the shape cuts. $\Theta$ is the angular
distance from the centre of the FoV. The initial event counting rate
$R_{(\gamma,cr)}$ can be calculated as
\begin{equation}
R_{(\gamma,cr)}(\Theta) = N_{(\gamma,cr)} \Delta t^{-1} \Delta \Omega^{-1},
\end{equation}
where $N_{(\gamma,cr)}$ is a number of registered events during the
time interval $\Delta t$ within the solid angle $\Delta \Omega$,
which is usually chosen to reflect the angular resolution of the
instrument. In calculations we used the cosmic-ray energy spectra
given in \cite{wibel}.

For diffuse and extended $\gamma$-ray sources there is no direct
correlation between the impact distance of a shower (the distance of
the shower axis to the geometrical centre of the array) and the
location of the image centroid position in the FoV. Showers detected
at the same impact distance but of different inclinations with
respect to the telescope axis could produce an image positioned
anywhere in the FoV, which is limited by the angular size of the
telescope imaging camera.

The calculated $\gamma$-ray detection rate, $R_{\gamma}$, as a
function of the angular distance from the telescope axis is shown in
figure~1 for different trigger multiplicities (the number of
triggered system telescopes). One finds a constant detection rate up
to $\Theta = 1^\circ$ (within 10\%) from the centre of the FoV and a
rather sharp decrease beyond that region. For the higher trigger
multiplicities, the fall-off in the detection rate starts at smaller
angular distance from the telescope axis.
% Figure 1 -----------------------------------------------------------
\begin{figure}[htbp]
\begin{center}
\includegraphics[width=0.6\linewidth]{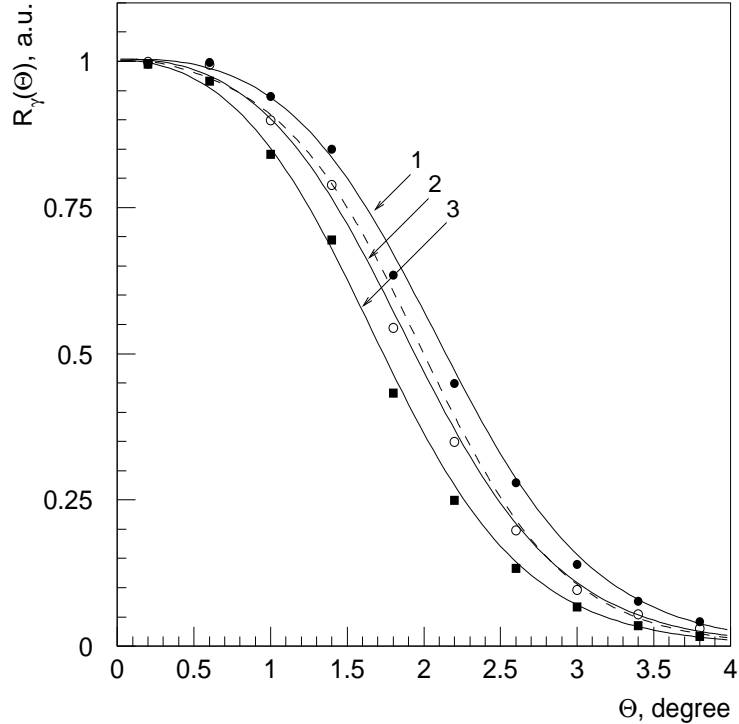}
\caption{{\em Detection rate of the isotropically distributed
$\gamma$-rays and cosmic rays calculated for the HEGRA system of
IACTs. The solid curves represent the $\gamma$-rays. Curves 1, 2 and
3 correspond to the trigger multiplicities of 2, 3 and 4,
respectively. The symbols correspond to the Monte Carlo data points
used in the fits. The dashed curve shows the rate of the cosmic-ray
showers for the trigger multiplicity of 2. Detection rates are
normalized to 1 at the centre of the FoV. The rates were calculated
after applying the quality cuts, distance $<1.8^\circ$, and image
size $>$40~photoelectrons. Hereafter, a.u.\ as indicated along the
$Y$-axis denotes arbitrary units.}}
\end{center}
\end{figure}
% --------------------------------------------------------------------
This rather sharp decrease can be explained as follows. In a toy
model one can assume that most of the detected $\gamma$-ray showers
have an impact distance around $R_o \simeq 120$~m (see, for example,
\cite{array}). The $\gamma$-ray-induced atmospheric showers, which
have a primary energy above the energy threshold of the instrument
($E > 500$~GeV), and which have shower axes directed along the joint
optical axis of the system telescopes, will produce the \v{C}erenkov
light images with the centroid positions shifted from the camera
centre by approximately $\Theta_o \simeq 1^\circ$. These images are
well within the effective radius of the HEGRA cameras of
$2.15^\circ$. Diffuse $\gamma$-rays, coming at the inclination angle
of about $\Theta_s \geq 1.1^\circ$ with respect to the telescope
axis, will often be truncated by the camera edge (see figure~2), and
the acceptance of these showers will be substantially lower due to
this effect. From such a toy model, one can expect a rather stable
$\gamma$-ray rate, $R_{(\gamma,cr)}(\Theta)$, up to $\Theta \simeq
1.1^\circ$. Note that, by increasing the energy of the detected
$\gamma$-ray showers, the effective impact distance increases as well
and consequently this effect of the camera edge shows up at smaller
shower inclinations.

% Figure 2 -----------------------------------------------------------
\begin{figure}[htbp]
\begin{center}
\includegraphics[width=1.\linewidth]{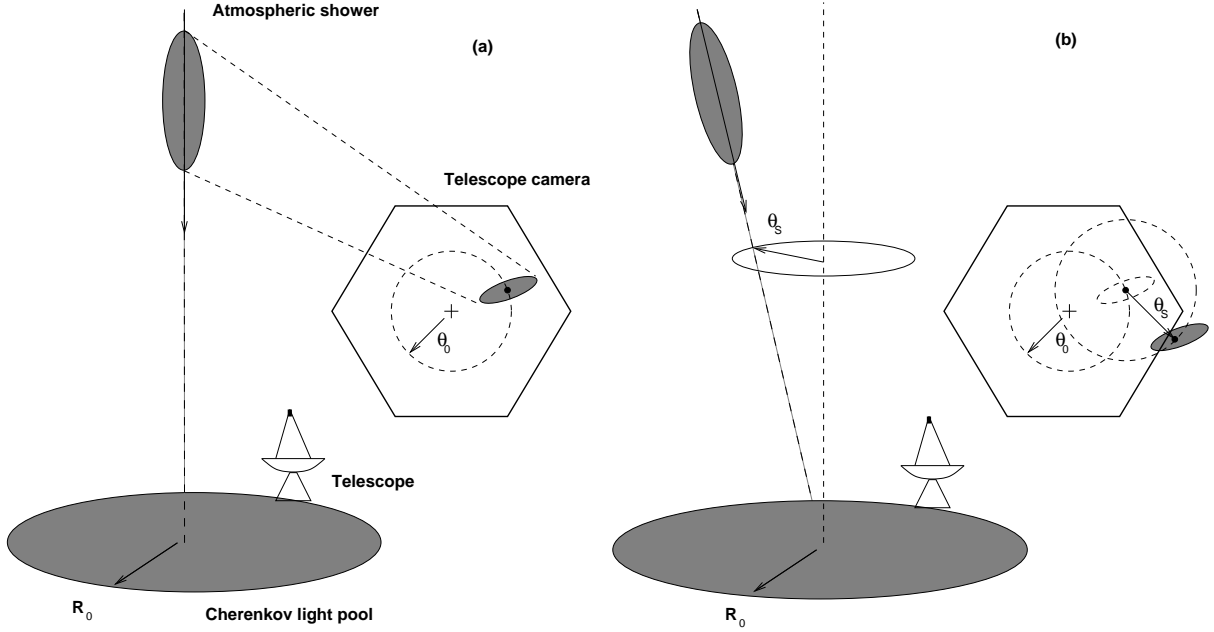}
\vspace{1mm} \caption{{\em Schematic diagram of the image
displacement in the camera focal plane.}}
\end{center}
\end{figure}
% --------------------------------------------------------------------

The detection rate as a function of the angular distance from the
centre of the FoV can be well fitted as
\begin{equation}
R_{(\gamma,cr)}(\Theta) = \rm a_0 \cdot (1+\Theta)^{a_1}[1+exp(a_2 \cdot (\Theta - a_3))]^{-1}.
\label{fit}
\end{equation}
The parameters of the fit are given in table 2 for the
$\gamma$-ray-induced and cosmic-ray-induced showers (see figure~1).

% Table 2 ------------------------------------------------------------
\begin{table}[htbp]
\caption{Parameters of the fit (equation~(\ref{fit})) for the
 simulated diffuse $\gamma$-ray-induced and cosmic-ray-induced
 atmospheric showers, registered with the HEGRA system of IACTs.}
\vspace{1 mm}
\begin{center}
\begin{tabular}{lccccc}
         & No of telescopes & $a_0$ & $a_1$ & $a_2$ & $a_3$ \\ \hline
         & 2          & 1.021 & 0.060 & 1.905 & 2.052 \\
$\gamma$ & 3          & 1.032 & 0.065 & 1.932 & 1.840 \\
         & 4          & 1.042 & 0.070 & 1.985 & 1.624 \\ \hline
         & 2          & 1.010 & 0.060 & 2.316 & 1.824 \\
cr       & 3          & 1.035 & 0.065 & 2.261 & 1.507 \\
         & 4          & 1.060 & 0.075 & 2.406 & 1.174 \\ \hline
\end{tabular}
\end{center}
\end{table}
% --------------------------------------------------------------------

\subsection{Angular resolution}
The angular resolution of the IACT array heavily relies on the
accuracy of the reconstruction of the image major axis, which in turn
depends mainly on the image size and the distance of the shower axis
to the centre of the array (impact distance). The images taken at
relatively large impact distances have an elongated angular shape and
offer an accurate determination of the image orientation. Note that
at very large impact distances (only for a central point source) the
images can be truncated by the camera edge. These images may distort
the reconstruction, but they can be removed from the analysis by
using the cut on angular distance of the image centre of gravity
(centroid) to the camera centre (here we use a cut on distance $<
1.8^\circ$). Thus, images with a rather large number of
photoelectrons (well above 40), and which are detected at impact
distances within the range of 50--200~m, offer the best determination
of the direction of the arriving shower \cite{HEGRA1}. This explains
why, with the increase in primary energy of the shower, the angular
resolution improves so much (for detailed discussions, see
\cite{HEGRA1,hofmann1}).

Finally, the angular resolution of the IACT array strongly depends on
the multiplicity of the system trigger \cite{HEGRA1}. The analysis of
the HEGRA Mkn~501 data sample \cite{hofmann1} has shown that using
three images out of four allows us to achieve an angular resolution
which is by roughly two times better than the resolution achieved
when using two images. At the same time, the use of higher trigger
multiplicities leads to a slightly higher energy threshold of the
instrument. All this may constrain the choice of the optimum analysis
scheme.

The angular resolution for a $\gamma$-ray source placed at a
different angular distance from the centre of the FoV is illustrated
in figure~3 and table~3. One can see that the geometrical
reconstruction of the direction of the arriving shower also works
very well for slightly inclined $\gamma$-ray showers, and the angular
resolution almost does not depend on the angle $\Theta$ within the
range of $\Theta \leq 1.5^\circ$. However, beyond this region the
images detected with the telescope system are often truncated by the
camera edge, which worsens the angular reconstruction. For the
precise measurements of an angular extension of a $\gamma$-ray source
one should take into account the actual angular resolution at a
certain angular distance from the centre of the FoV, and perform the
deconvolution to obtain the measured angular extension of the source.

% Figure 3 -----------------------------------------------------------
\begin{figure}[htbp]
\begin{center}
\includegraphics[width=0.6\linewidth]{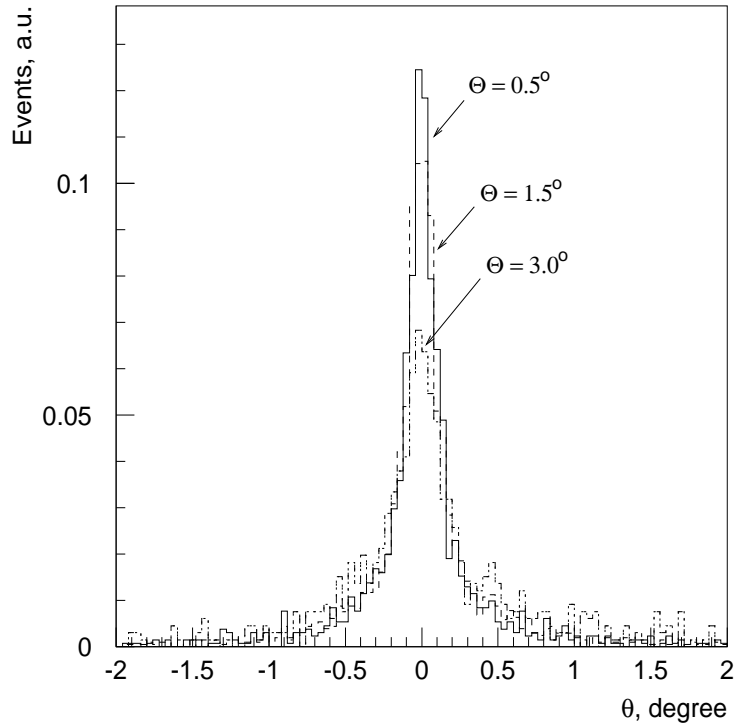}
\caption{{\em The angular resolution of the OFF-axis $\gamma$-ray showers
  detected with the HEGRA system of IACTs. At least three images were
  used for the reconstruction of the direction of the arriving shower.}}
\end{center}
\end{figure}
% --------------------------------------------------------------------

% Table 3 ------------------------------------------------------------
\begin{table}[htbp]
\caption{Angular resolution for an individual off-axis
  $\gamma$-ray-induced atmospheric shower ($\Delta \theta$ defines a 68\%
    error circle around an actual source position). }
\begin{center}
\begin{tabular}{lllllll}
$\Theta$ ($^\circ$)        & 0.5  & 1.0  & 1.5  & 2.0  & 2.5  & 3.0 \\ \hline
$\Delta \theta$ ($^\circ$) & 0.12 & 0.12 & 0.14 & 0.16 & 0.17 & 0.31 \\ \hline
\end{tabular}
\end{center}
\end{table}
% --------------------------------------------------------------------

\subsection{Rejection of cosmic rays}
Apart from the excellent angular resolution, the IACT system offers
the ability to reject cosmic-ray showers using the shape of the
\v{C}erenkov light images registered in a number of telescopes.
Imaging cameras detect, in addition to the \v{C}erenkov light from
the atmospheric showers, all kinds of background light (e.g.~night
sky background light, direct star light, etc). To remove most of the
camera pixels containing exclusively the background light, a specific
image clearing procedure is usually applied. The pixels with light
intensity below some fixed intensity limit are sorted out of the
image. The images of the high-energy $\gamma$-ray showers detected
close to the telescope system embrace many more pixels as compared to
low-energy events or to the same event but at rather large impact
distance. Using the standard second moment analysis, one can obtain
very different angular sizes of recorded images, depending on the
primary shower energy and shower impact distance.

% Table 4 ------------------------------------------------------------
\begin{table}[t]
\caption{Acceptances of $\gamma$-ray-induced and cosmic-ray-induced
  atmospheric showers after the analysis by a mean scaled width
  cut of 1.1.}
\vspace{0.5mm}
\begin{center}
\begin{tabular}{llllllll}
Trigger    & $\Theta$ ($^\circ$) & 0.5  & 1.0  & 1.5  & 2.0  & 2.5  & 3.0 \\ \hline
  2/5      & $k_{\gamma}$    & 0.67 & 0.68 & 0.69 & 0.62 & 0.58 & 0.41 \\
  3/5      & $k_{\gamma}$    & 0.66 & 0.67 & 0.68 & 0.62 & 0.55 & 0.34
  \\
  4/5      & $k_{\gamma}$    & 0.65 & 0.65 & 0.66 & 0.60 & 0.50 & 0.21
 \\ \hline
  2/5      & $k_{CR}$        & 0.06 & 0.06 & 0.06 & 0.06 & 0.07 & 0.08\\
  3/5      & $k_{CR}$        & 0.04 & 0.04 & 0.04 & 0.04 & 0.05 &
  0.07 \\
  4/5      & $k_{CR}$        & 0.03 & 0.03 & 0.03 & 0.04 & 0.06 & 0.08
  \\ \hline
\end{tabular}
\end{center}
\end{table}
% --------------------------------------------------------------------

In order to avoid the dependence of an image on the shower energy and
impact distance, one can derive using the standard image parameter
width, the parameter mean scaled width, $\langle \tilde w
\rangle$~\cite{padova,kruger} for each individual event:
\begin{equation}
\langle \tilde w \rangle = 1/n \sum_{k=1}^n w_k / \langle w
\rangle_k^{ij}.
\end{equation}
Here $n$ is the number of triggered telescopes, $w_k$ is the width
for the $k$th image and $\langle w \rangle_k^{ij}$ is the expected
mean image width calculated beforehand from the Monte Carlo
simulations over a number of bins on the impact distance ($\Delta
r_i, i=1,n$) and image size ($\Delta log(S_j), j=1,m$) \cite{HEGRA1}.
The mean scaled width parameter has been proven to be an effective
tool in rejection of a substantial fraction of the cosmic-ray
showers. For central point-like $\gamma$-ray sources, applying a cut
on mean scaled width of $\langle \tilde w \rangle < 1.1$ provides for
the system trigger 4/4 an acceptance of $\gamma$-rays at 65\%,
whereas the cosmic-ray contamination is reduced by this cut by a
factor of 30, corresponding to an enhancement of the $\gamma$-ray
sample (Q-factor = $k_\gamma k_{CR}^{-1/2}$) of 3.7.

% Figure 4 -----------------------------------------------------------
\begin{figure}[htbp]
\begin{center}
\vspace{5mm}
\includegraphics[width=0.6\linewidth]{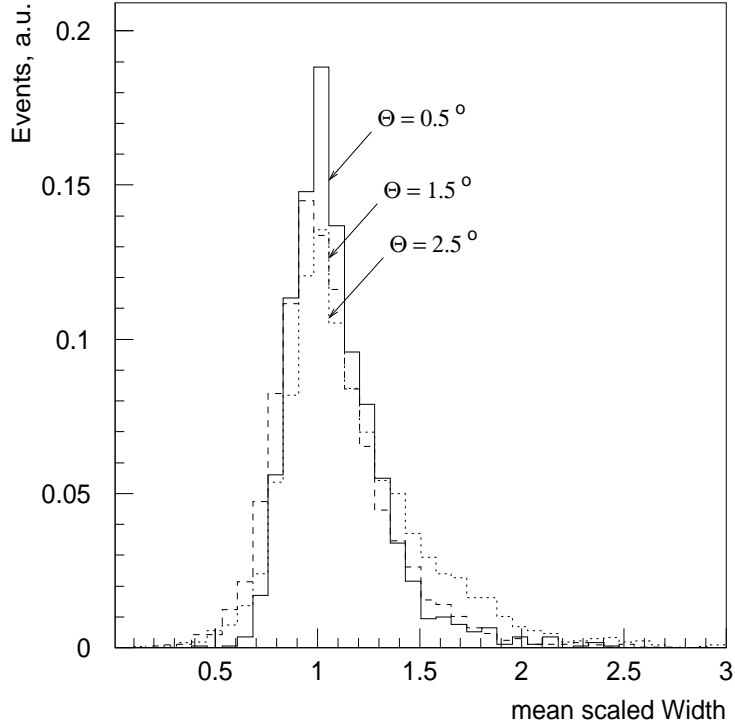}
\caption{{\em The distribution of the mean scaled width parameter for
 the samples of $\gamma$-rays at three inclination angles with respect
 to the telescope optical axis. Events with at least two images were
 used in the analysis.}}
\vspace{2mm}
\end{center}
\end{figure}
% --------------------------------------------------------------------

Here we study how the cut on mean scaled width works for the
$\gamma$-rays coming off the telescope optical axis. The calculations
have been made for six bands, each of a size of 0.2$^\circ$, centred
at the angular distances of $0.5^\circ$, $1^\circ$, $1.5^\circ$,
$2.0^\circ$, $2.5^\circ$ and $3.0^\circ$, from the centre of the FoV.
The distributions of the mean scaled width for the $\gamma$-ray
showers within three of these bands are compared in figure~4. With an
increasing angle relative to the axis, the distributions remain
peaked at 1, but develop a tail towards larger $\langle \tilde w
\rangle$. The acceptances of the $\gamma$-rays after applying the cut
on mean scaled width are almost identical for these bands. We may
conclude that the shape of the images for the $\gamma$-ray showers
coming off the telescope optical axis at angles less than $\simeq
1.5^\circ$ are still not noticeably affected by the telescope camera
edge. One can see a slight decrease in the $\gamma$-ray acceptance,
$k_\gamma$, towards larger displacements from the centre of the FoV,
which is related to the effect of the camera edge. For the
isotropically distributed cosmic rays, the acceptance after applying
the mean scaled width cut ($k_{cr}$) also does not depend on the
angular distance within the range of $\Theta \leq 2^\circ$ (see
table~4), whereas it increases slightly towards larger angular
distances.

\section{Simulations versus data}
The simulations of the diffuse $\gamma$-rays can be verified by the
observations of the Crab Nebula. The observations in the wobble mode
correspond to the angular displacement of the source by $0.5^\circ$
from the centre of the FoV. Assuming in the simulations the Crab
Nebula flux and spectrum as measured by the HEGRA collaboration
\cite{hegra_crab}, one can calculate the corresponding $\gamma$-ray
rate at the same angular distance from the centre of the FoV. The
results for different trigger multiplicities are summarized in
table~5. The Monte Carlo simulations reproduce well the relative
rates for two, three and four telescopes. Note that the possible
systematic error of the Monte Carlo simulated rates is estimated to
be less than 10\%.

The distributions of the mean scaled width parameter for the
$\gamma$-ray-induced and cosmic-ray-induced showers are shown in
figure~5. The Monte Carlo distributions fit very well with the data.
Here, the same Monte Carlo tables of mean width as function of impact
distance and image size have been used for the simulated and recorded
events. For each individual event, the Monte Carlo tables were
interpolated according to the reconstructed impact parameter and
measured image size in simulations and in the data. Simulations are
in very good agreement with the data. The position of the peak of the
simulated $\gamma$-ray showers agrees within 3\% with the data. The
data distribution of the $\gamma$-rays was produced using the HEGRA
Crab Nebula data sample \cite{hegra_crab}.

% Table 5 ------------------------------------------------------------
\begin{table}[t]
\caption{The calculated and measured detection rate,
  $R_{\gamma},\,hr^{-1}$, of the Crab Nebula at 0.5$^\circ$
  distance from the centre of the FoV.}
\vspace{2mm}
\begin{center}
\begin{tabular}{llll}
Trigger     & 2/4      & 3/4      & 4/4 \\ \hline
Data        & 18$\pm$2 & 21$\pm$3 & 19$\pm$3 \\
Monte Carlo & 20       & 20       & 22    \\ \hline
\end{tabular}
\end{center}
\vspace{0.5mm}
\end{table}
% --------------------------------------------------------------------

% Figure 5 -----------------------------------------------------------
\begin{figure}[htbp]
\begin{center}
\includegraphics[width=0.6\linewidth]{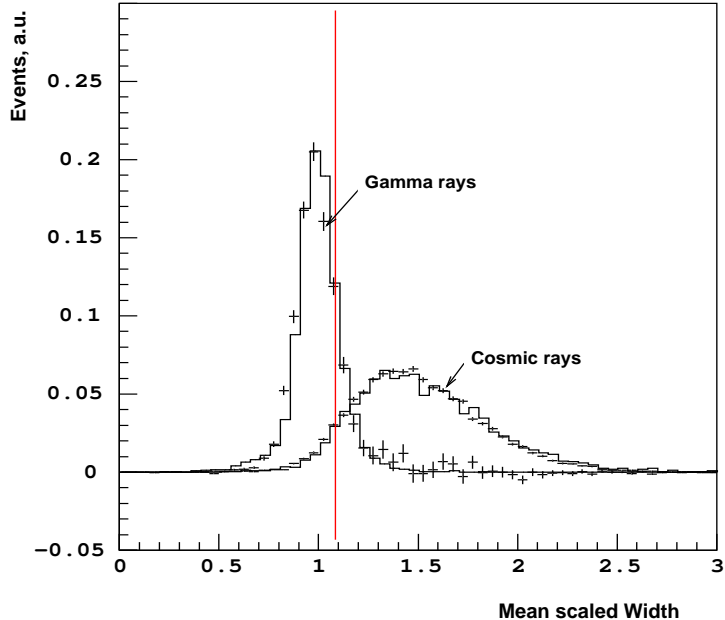}
\caption{\em The distributions of the mean scaled width parameter for
 the sample of $\gamma$-rays and cosmic rays. Histograms show the Monte
  Carlo simulations. The crosses denote the HEGRA data taken with the
  system trigger as two telescopes out of five. The distributions are normalized
  to one particle. The content of the histogram bins is given in arbitrary
  units.}
\end{center}
\end{figure}
% --------------------------------------------------------------------

\section{Sensitivity over FoV}
The sensitivity of the HEGRA system over the whole FoV can be
estimated using the detection rates and acceptances previously
discussed. To parametrize the sensitivity of the instrument one can
use the $Q$-parameter calculated as
\begin{equation}
Q(\Theta) = k_{\gamma} R_{\gamma}
\cdot [k_{CR} R_{CR}]^{-1/2}
\label{qf}
\end{equation}
where the rates of the $\gamma$-rays and cosmic rays are normalized
to 1 at $\Theta = 0$. The results of such calculations are shown in
figure~6. One can see that the value of the $Q$-parameter is constant
up to $\Theta = 1^\circ$. Even though the $\gamma$-ray detection rate
decreases noticeably beyond that region the $Q$-parameter at $\Theta
= 1.5^\circ$ is only 5\% less than in the central part of the FoV.

The higher telescope multiplicities provide a better Q-factor. This
is due to the fact that for higher multiplicities the cosmic-ray
rejection is substantially better, even though the detection rate of
the $\gamma$-rays decreases at larger angular distances from the
centre of the camera. The sharp decrease in $\gamma$-ray rate at
large angular distances ($\Theta\geq 2.5^\circ$) finally makes
observations with higher telescope multiplicities less effective.

Note that the angular resolution is
almost constant up to the angular distance of $1^\circ$.

% Figure 6 -----------------------------------------------------------
\begin{figure}[t]
\begin{center}
\includegraphics[width=0.6\linewidth]{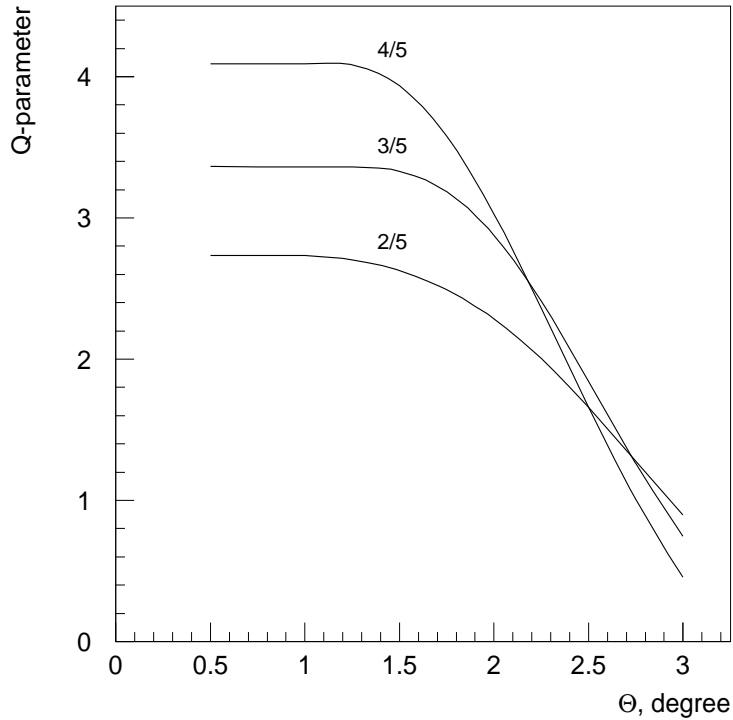}
\caption{\em Q-factor (equation~(\ref{qf})) calculated at
  different angular distances from the centre of the FoV. Corresponding trigger
  multiplicities are shown.}
\end{center}
\end{figure}
% --------------------------------------------------------------------

In observations of the Crab Nebula with the HEGRA system of IACTs at
zenith angles $\leq 35^\circ$, the numbers of $\gamma$-ray and
cosmic-ray showers detected in one angular bin ($\Delta \Omega \simeq
5 \times 10^{-5}$~str), placed at the centre of the camera, in one
hour (before applying the shape cuts) are about 80 and 406 ($\rm
counts\,hr^{-1}$), respectively. These measured rates can be
reproduced with high accuracy using the Crab Nebula spectrum as
measured by HEGRA \cite{hegra_crab}, and using the fluxes and spectra
of cosmic rays given in \cite{wibel}. One can calculate the
signal-to-noise ratio after 1~h observations as $\rm S/N = (ON -
OFF)/(ON+OFF)^{1/2}= \,N_\gamma/[(N_\gamma+N_{cr})+N_{cr}]^{-1/2}
\cdot Q(\Theta)$. Thus, in the centre of the FoV the signal-to-noise
ratio is $\rm S/N \simeq 7.5\sigma\, hr^{-1}$ for the trigger
multiplicity 2/5, which agrees with the value directly measured in
observations of Crab Nebula ($\rm S/N \simeq 7.6\sigma\, hr^{-1}$).
One can obtain an improvement by a factor of 1.25 and 1.5 for the
multiplicities of 3/5 and 4/5, respectively, on the Crab Nebula data
sample as well as for the simulated data.

\section{Estimate of the significance of excess}
In observations of extended $\gamma$-ray sources or diffuse
$\gamma$-ray emission from certain regions of the sky, it is not
trivial to calculate properly the significance of a possible excess
of the $\gamma$-ray candidates. In order to calculate this
significance, one could follow different procedures, depending on
whether the OFF-runs (data taken out of the source direction) are
available or not. However, in both cases it is necessary to calculate
the detection efficiency versus the radial distance from the centre
of the FoV directly from the data and, afterwards, to apply a flat
fielding in order to compensate for the non-uniform response of the
camera at different angular distances. One can fit the detection rate
of the cosmic rays versus the angular shift from the centre of the
FoV and then apply that fit to all angular bins over the entire FoV.
Such a fit can vary accordingly to the source zenith angle and the
night sky background, and can also have an azimuthal dependency.
However, our studies of the azimuth dependence did not reveal a
significant effect for the rather symmetric HEGRA system of IACTs.

Since the stereoscopic reconstruction of atmospheric showers with the
HEGRA system of IACTs allows us to calculate the right ascension (RA)
and declination (DEC) for each individual event as well as the
angular slopes of the shower axis in the joint focal plane, it is
possible to produce a two-dimensional map of the reconstructed events
over the FoV, after applying the standard cut on mean scaled width,
e.g.~$\langle \tilde w \rangle<$1.1, in order to separate
$\gamma$-rays from hadrons. The observational window can be divided
into square cells of size $0.2^\circ \times 0.2^\circ$, which is
close to the optimum, given the angular resolution of the HEGRA
system of IACTs (RMS$\simeq 0.1^\circ$). At the same time such bin
size allows us to have a relatively large number of pixels for
computing the average number of cosmic-ray hits per angular bin, which can be
used to estimate the average background content per angular bin. As
noted in section~3, the region of almost constant sensitivity is
limited by roughly 1$^\circ$ from the centre of the FoV. Given the
number of counts for each bin, one can calculate the significance for
each bin, $S$, using the corresponding OFF data sample.

Let us now examine two cases separately. If both ON-source and
OFF-source data are available, after applying the mean scaled width
cut the mean number of counts per angular bin in ON-source and
OFF-source data samples can be calculated as
\begin{equation}
\langle N_{(on,off)} \rangle=\frac {1}{M}\sum_{i=1}^{M}
  N_{(on,off)}(i,\Theta) f^{-1}(\Theta)
\label{on}
\end{equation}
where $N_{(on,off)}(i,\Theta)$ is the number of counts in $i$th bin
at a certain distance, $\Theta$, from the centre of the FoV. Function
$f(\Theta)$ accounts for the non-constant event rate at different
radial distances. $M$ is the total number of bins in the grid.
$\langle N_{off} \rangle$ is the mean number of counts per bin placed
at the centre of the camera's FoV in the OFF-source sample. Given
this number, one can estimate the background content for each camera
pixel ({\it i}) for ON-source data as
\begin{equation}
N_{off}^{(1)}(i) = \alpha f(\Theta) \langle N_{off} \rangle
\end{equation}
The scaling factor $\alpha$ can be determined as
\begin{equation}
\alpha=  {\langle N_{on} \rangle} {(\langle N_{off} \rangle)}^{-1},
\end{equation}
which equalizes the difference in observational time for ON-source
and OFF-source samples, as well as possible differences in
corresponding zenith angle distributions.

In the case when only ON-source data are available, one can use a
mean number of counts per bin\footnote{Note that in observations of
extended $\gamma$-ray sources with well-defined location and
extension one can exclude from the averaging loop
(equation~(\ref{on})) the angular bins, which are {\it a priori}
covered by the source. However, the source extension has to be
significantly smaller than the observational window used.} averaged
over all bins in the ON-source data sample, $\langle N_{on} \rangle$.
It is assumed  here that the expected rate of the $\gamma$-rays is
very low and is negligible compared with the cosmic-ray rate.

Thus, the mean number of predicted  background counts in each angular
bin is given by
\begin{equation}
N_{off}^{(2)}(i) = \alpha f(\Theta) \langle N_{on} \rangle.
\label{on1}
\end{equation}
Here the scaling factor $\alpha$ is calculated  as $\alpha \simeq
1/M$, where $M$ is a total number of bins used in
equation~(\ref{on}).

Finally, the significance for each individual angular bin can be
estimated using the approach suggested in \cite{li_ma}. Assuming that
the bins with high signal-to-noise ratio ($\simeq3.5\sigma$) contain
an excess of $\gamma$-rays one can exclude these pixels from
calculating the average number of background counts per bin and
recalculate the significances again. Such an iterative procedure can
be repeated a number of times before the values of the significances
finally converge.

The background model constructed from the ON-source data itself
cannot be applied for observations of a presumably diffuse
$\gamma$-ray source of a large angular size, which is comparable with
the angular size of the entire FoV of the instrument. However it
might be effective in a discovery mode while scanning very extended
sky regions.

For a positive detection of a quite extended $\gamma$-ray source
(with a size larger than $0.5^\circ$), one can expect that a few
contiguous pixels should show a high value of significance. One can
calculate the confidence level of the excess using the approach
suggested by Li and Ma \cite{li_ma} for such a particular case. This
approach takes into account that the confidence level of an excess in
each bin depends, in addition, on the total number of bins (number of
trials) $M$, which may be considered as $M$ independent measurements.

After the average background calculation, we can calculate the
standard deviation for each pixel from the average background.
Naturally, only those pixels which show an excess in the ON data
sample, (for instance, with $S>3.5\sigma$, where $\sigma$ is one
standard deviation for the Gaussian distributed background events),
might be considered for statistical analysis. The probability $p$ to
achieve 3.5$\sigma$ excess due to fluctuations in background for each
individual pixel, can be simply calculated as
\begin{equation}
  p = 1 - \frac{1}{\sqrt{2\pi}} \int_{-\infty}^{3.5} e^{-\frac{1}{2}t^2} dt.
\end{equation}
Eventually, the probability to obtain $k$ pixels within our FoV will
be given by
\begin{equation}
  p_k = C_M^k p^k (1-p)^{M-k}
  \label{ppp}
\end{equation}
($C_M^k$ is the binomial coefficient). The results of calculations
for different pixel multiplicities in excess and $M=80$ using
equation~(\ref{ppp}) are provided in table~\ref{t1}.
% ----- Table 1 ------------------------------------------------------
\begin{table}[t]
\caption{Estimates of the confidence level for the excess events. $k$
  is the number of pixels showing a $3.5\sigma$ excess from the
  average background.}
\begin{center}
\begin{tabular}{llll}
$k$ & $p_k$ Eqn.(\ref{ppp}) & $p_k$ MC \\ \hline
 1 & 1.83$\cdot 10^{-2}$ & 1.83$\cdot 10^{-2}$ \\
 2 & 1.68$\cdot 10^{-4}$ & 1.70$\cdot 10^{-4}$ \\
 3 & 1.02$\cdot 10^{-6}$ & 9.60$\cdot 10^{-7}$ \\
 4 & 4.55$\cdot 10^{-9}$ & 4.70$\cdot 10^{-9}$ \\ \hline
\end{tabular}
\end{center}
\label{t1}
\end{table}
% --------------------------------------------------------------------
The calculations show, for example, that a four-pixel coincidence
within the observational window composed of 80 pixels is an extremely
rare case with a probability less than $10^{-8}$ and the
corresponding estimate of confidence at 6$\sigma$ level. But, in
practice, one could observe one, two, three or even many more bins
with a significant excess in the event rate. All these cases should
be statistically tested taking into account the total number of
selected angular bins within the FoV, which corresponds to the number
of independent trials.

In order to prove the formulae given by equation~(\ref{ppp}), we have
also made straightforward Monte Carlo simulations, randomizing the
number of entries for $M=80$ pixels according to a Gaussian
distribution with average value $\langle N_{ON}(i) \rangle = 300$.
The results are summarized in table~\ref{t1} (see column $p_k$ MC).
One can see a rather good agreement between the direct Monte Carlo
simulations and the values obtained from equation~(\ref{ppp}), taking
into account a limited statistics in the Monte Carlo simulations
which finally limits the accuracy of the calculations.

An alternative approach to estimate the significance for extended
sources would be to make some assumptions about the position and the
extension of the $\gamma$-ray source, based on physical arguments,
taking as ON, the correspondent sky region, and as OFF, several sky
regions of the same shape (if possible). One can calculate the
corresponding significance using the approach of Li and Ma
\cite{li_ma}.

\section{Sensitivity to extended $\gamma$-ray sources}

TeV $\gamma$-ray emission, resulting from the $\pi^\circ$-decay
produced in nearby supernova remnants (SNR) by the accelerated
protons colliding with the ambient thermal gas nuclei, is marginally
high enough to be detectable by currently operating imaging
atmospheric \v{C}erenkov telescopes (see, for example,
\cite{bv97,bkv02}). Given the angular size of the shell-type SNRs,
which has been measured in radio wavelength range, the emission
region in TeV $\gamma$-rays is expected to be significantly extended
\cite{bkv02}. In good approximation, the source is characterized by
the constant brightness in $\gamma$-rays all over the SNR radio
extent.

% Figure 8 -----------------------------------------------------------
\begin{figure}[htbp]
\begin{center}
\includegraphics[width=0.45\linewidth]{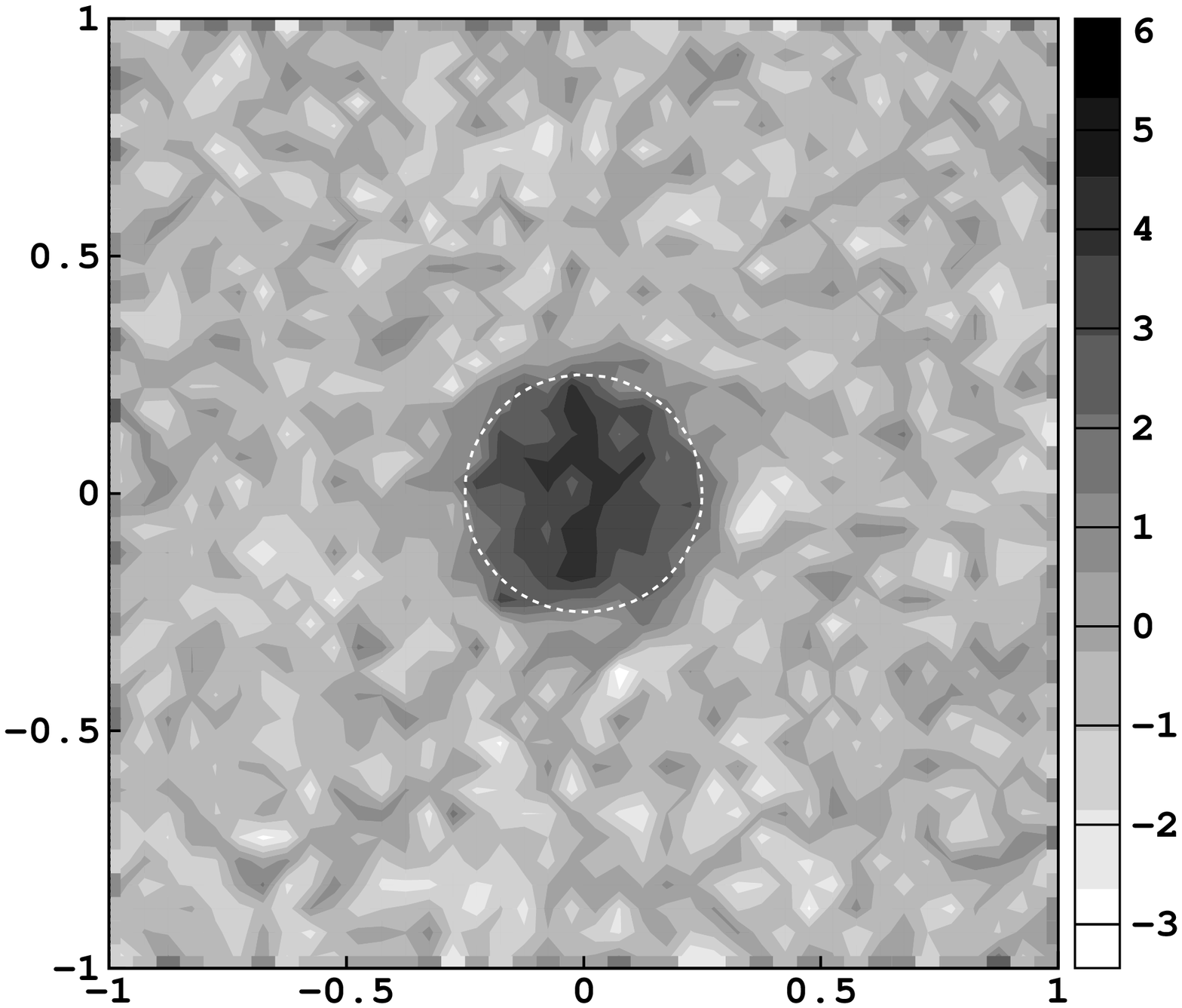}\\
\includegraphics[width=0.45\linewidth]{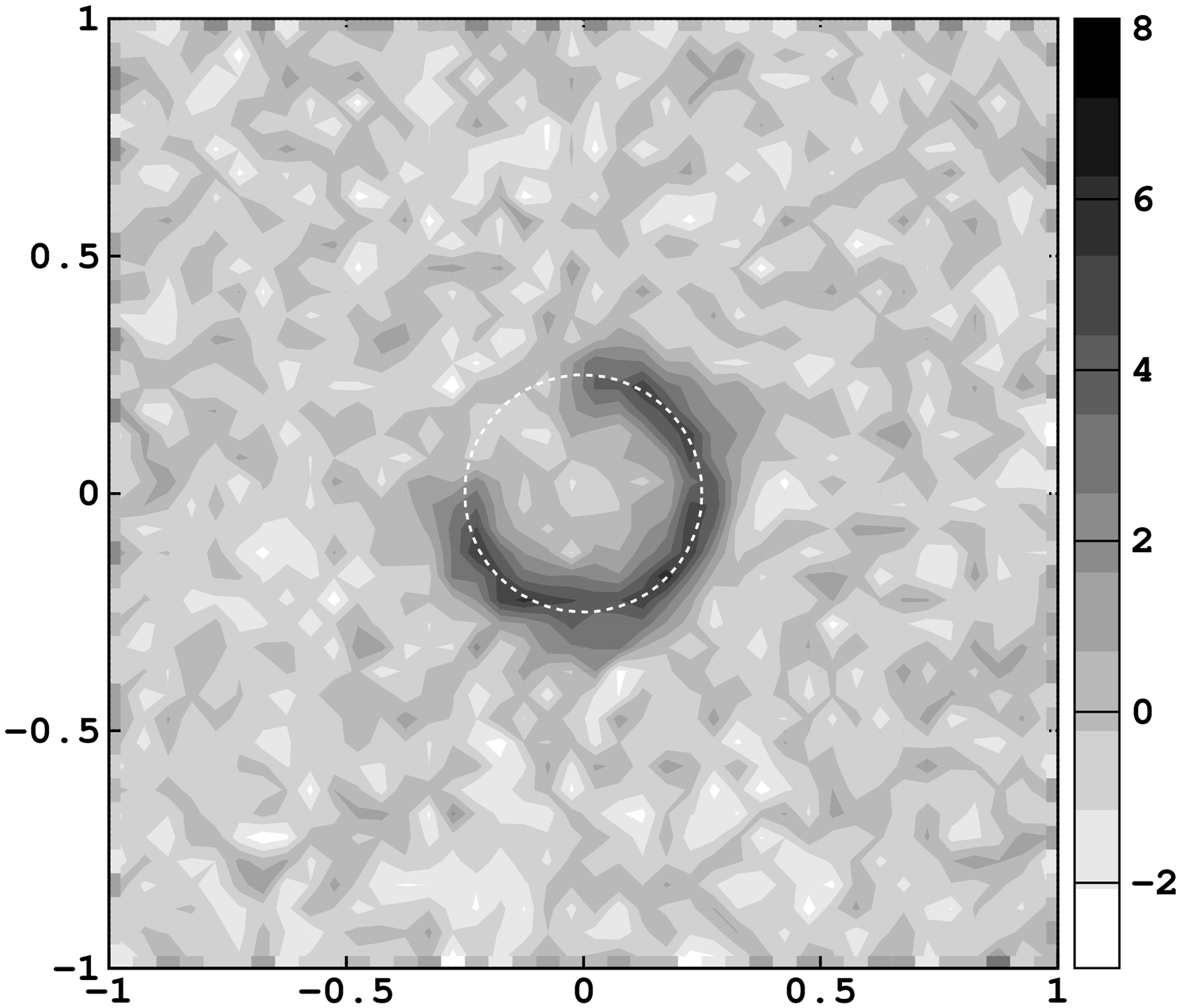}\\
\includegraphics[width=0.45\linewidth]{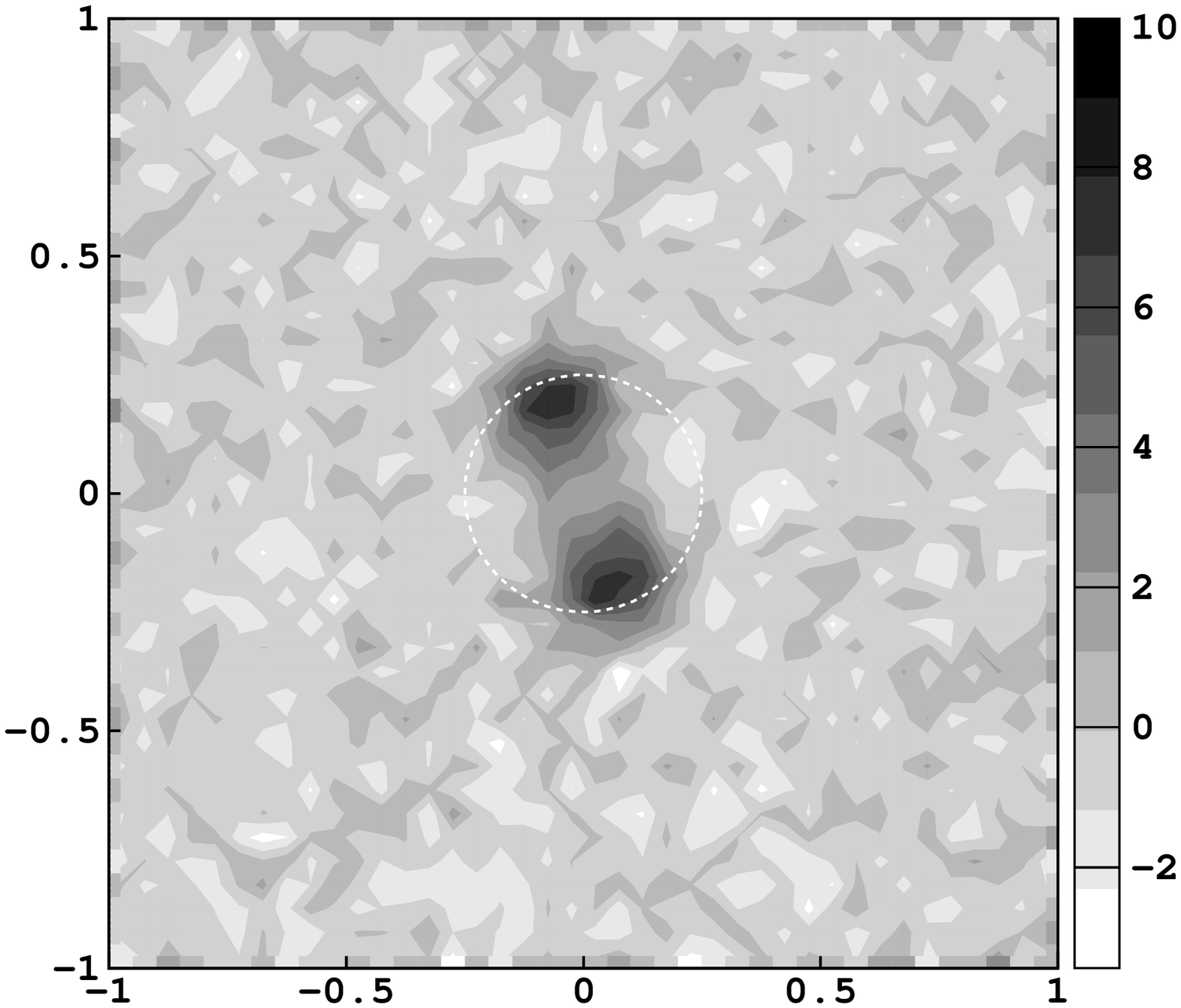}
\vspace{1mm} \caption{{\em Monte Carlo simulated response of the
HEGRA system of five IACTs (Trigger: 2/5) after 100~h of observations
of shell-type SNRs for three models of TeV $\gamma$-ray emission:
`circle' (upper panel); `ring' model for the $\gamma$-rays simulated
over 75\% of the entire ring (central panel); `double-pole' model
(lower panel). For details on radial profiles of TeV $\gamma$-ray
emission see \cite{bkv02}. The angular size of the SNR in radio
wavelength is indicated by the dashed circle.}}
\end{center}
\end{figure}
% --------------------------------------------------------------------

Based on the detailed Monte Carlo simulations for the HEGRA system we
have estimated the sensitivity of the instrument in observations of
extended $\gamma$-ray sources. The results of the simulations can be
interpreted using the established sensitivity for a point-like
$\gamma$-ray source. Thus, the resulting signal-to-noise ratio for a
$\gamma$-ray source of an arbitrary angular size, $\theta$, can be
calculated as
\begin{equation}
S/N = \tilde R_\gamma / [\tilde R_{CR} \cdot (\theta/\theta_0)^2]^{1/2}
t^{1/2} =
\tilde R_\gamma /[\tilde R_{CR}]^{1/2} \cdot (\theta/\theta_0)^{-1} t^{1/2}
\end{equation}
where $\tilde R_\gamma$ and $\tilde R_{CR}$ are the detection rates
of $\gamma$-rays and cosmic rays, respectively, after applying the
analysis cut. Values $\theta_0$ and $\theta$ correspond to the
orientational cuts for a point-like source ($\theta_0$= 0.1) and
extended $\gamma$-ray source ($\theta$ is a source extension).
$\tilde R_{CR}$ is a cosmic-ray rate for a standard orientational cut
applied for a point source search. For the isotropic cosmic-ray
background, the total number of background counts for an extended
$\gamma$-ray source is proportional to the angular area covered by
this source ($R_{CR} \propto \theta^2$). For a 5~$\sigma$ detection
and fixed observational time\footnote{For instance, SNR Cas~A was
observed with HEGRA system of IACTs for a total exposure time of
250~h \cite{hegra_casa}.} ($t=100$~h) one can calculate the minimum
detectable flux from the point-like $\gamma$-ray source. For the
HEGRA array of five IACTs such a minimum detectable flux is
$F^{p.s.}_\gamma(\ge 1~TeV) = 1.1 \cdot 10^{-12}\rm
~photon~cm^{-2}~s^{-1}$. Finally, the minimum detectable flux for an
extended $\gamma$-ray source is given by
\begin{equation}
F^{ex.s.}_\gamma = F^{p.s.}_\gamma \cdot (\theta /\rm 0.1~deg),\label{ss}
\end{equation}
where $\theta$ is an angular radius of a source region. Note that for
an extended $\gamma$-ray source of large extension
($\theta>0.5$~degree) the integral rates $\tilde R_\gamma$ and
$\tilde R_{CR}$ need to be corrected for the fall-off in the camera
response function (see figure~1). For $\gamma$-ray sources of large
extension ($\theta>1$~degree) one has to perform observations
preferably in ON/OFF mode, i.e.~taking the same amount of OFF data as
for the ON data sample, because almost the entire sensitive area is
covered by a source region. The results of the detailed Monte Carlo
simulations may deviate from the estimates given by
equation~(\ref{ss}) for the $\gamma$-ray sources of angular extension
substantially larger than 1~$^\circ$.

As an example, we consider here a shell-type SNR with an angular size
of 0.25$^\circ$ (such as SN~1006). Three different models of TeV
$\gamma$-ray emission have been discussed in \cite{bkv02}. The
profiles of $\gamma$-ray brightness are taken from \cite{bkv02}. The
leptonic model of $\gamma$-ray emission yields the uniform
distribution of emitted $\gamma$-rays (`circle' model), whereas for
hadronic models the brightness might peak around the shock front
(`ring' model) or concentrate around two emission poles
(`double-pole' model) \cite{bkv02} (see figure~7). We have modelled
the response of the HEGRA system of five IACTs for the fluxes and
brightness profile given in \cite{bkv02} after 100~h of observations
(see figure~7) assuming the $\gamma$-ray flux given in
\cite{tanimori}. The minimal observing time for a 5~$\sigma$
detection depends on the assumed morphology of the $\gamma$-ray
emission. Thus, for the `circle', `ring' and `double-pole' models the
minimal detection times are 16~h, 5~h and 2~h, respectively. Deep
observation of such an SNR could allow us to study in detail the
morphology of the TeV $\gamma$-ray emission and finally to
distinguish between different models of $\gamma$-ray emission.

\section{Conclusion}

The response of the HEGRA system of IACTs to the diffuse and extended
$\gamma$-ray emission over the FoV of the instrument was studied by
means of detailed Monte Carlo simulations. Within the angular region
limited by 1$^\circ$ from the centre of the FoV the detection rate of
the $\gamma$-rays as well as the quality factor, characterizing the
efficacy of the $\gamma$-ray selection, are constant. Further
extension of the observational window up to 1.5$^\circ$ still allows
us to have the same sensitivity to the $\gamma$-ray fluxes but with
noticeably reduced $\gamma$-ray detection rate. An analysis of
different trigger multiplicities reveals an improvement in the
sensitivity, whereas higher multiplicities lead to the substantial
decrease in the $\gamma$-ray detection rate.

We have modelled the response of the HEGRA system of five IACTs for
observations of nearby SNRs. Even though the final sensitivity
estimate depends on the actual morphology of the TeV $\gamma$-ray
emitting region, in a simple case of a `circular' emission region the
sensitivity might be derived by rescaling the sensitivity for a
point-like source.

The studies discussed here could have a general use for the
forthcoming arrays of IACTs such as CANGAROO, H.E.S.S.\ and VERITAS.

\noindent {\it Acknowledgments.} This work was supported by CICYT
(Spain) and BMBF (Germany).

\end{document}